\documentstyle[12pt]{article}
\renewcommand{\deg}{^\circ}
\newcommand{\comment}[1]{}
\newcommand{\rip}[1]{\left| {#1} \right\rangle}
\newcommand{\trio}[3]{{\pil{ #1 \abs{ #2 } #3}}}

\newcommand{\pil}[1]{{\left\langle #1 \right\rangle}}
\newcommand{\pib}[1]{{\left[ #1 \right]}}
\newcommand{\pip}[1]{{\left( #1 \right)}}

\newcommand{\abs}[1]{{\left| #1 \right|}}
\newcommand{\w}[1]{{\mbox {#1}}}

\newcommand{\ov}[1]{{\overline {#1}}}
\newcommand{\eqnb}{\begin{equation}}
\newcommand{\eqne}{\end{equation}}
\newcommand{\sep}[1]{{\hspace{#1 in}}}
\newcommand{\smfrac}[2]{\textstyle {\frac{#1}{#2}}}
\newcommand{\ipt}[2]{\left\langle #1 | #2 \right\rangle}

\newcommand{\pub}[1]{{\it {#1}}}
\newcommand{\vol}[1]{{\bf {#1}}}

\newcommand{\unka}{\perp}
\title{Nucleon strangeness and spin crisis?} 
\author{L.~Babukhadia$^{a}$ and M.~D.~Scadron$^{b}$ \\\\\\ 
	$^{a}$ Department of Physics, State University of New York, \\
	Stony Brook, NY 11794 USA \\
	Electronic mail: blevan@fnal.gov \\\\
	$^{b}$ Department of Physics, University of Arizona,\\
	Tucson, AZ 85721 USA \\
	Electronic mail: scadron@physics.arizona.edu}
\date{}

\begin{document}

\maketitle
\bigskip
\begin{abstract}
We summarize three alternative extensions of the quark-valence picture 
of nucleon strangeness and spin, including (1) gluon spin in QCD,
(2) Bjorken sum rule and QCD, (3) $D$--$E$ meson mixing and tadpole
leakage.  All three approaches suggest that $\sim$ 40\% of the
nucleon's spin (as well as its momentum) resides with quarks,
but that less than 6\% is due to strange quarks.
\end{abstract}
\vfill
\thispagestyle{empty}

\newpage
\section*{I.  Quark Valence Picture}
	Given that pions are composed of $\ov{q}q$ nonstrange ($NS$)
quarks and (static) nucleons are composed of $qqq$ $NS$ quarks 
$uud$ and $ddu$, the strangeness-antistrangeness quantum numbers
appear when e.g.\ $\pi^-p \to \Lambda K^0$.  The static
nonrelativistic quark model then requires the SU(6)
valence ($V$) values for the axial-vector quark spin component
($\Delta q \sim \trio{N}{\ov{q}\gamma_\mu \gamma_5 q}{N} S^\mu$)
matrix elements of the nucleon to be
$$
	\Delta u_V = \smfrac{4}{3}~,\sep{.3}
	\Delta d_V = -\smfrac{1}{3}~,\sep{.3}
	\Delta s_V = 0~.
\eqno(1)
$$
This quark valence picture (1) was first tested using the Sachs form
factor for on-shell nucleons $\Gamma_\mu = G_E (q^2) P_\mu / m_N
+ G_M (q^2) r_\mu / 
4m^2_N$, where $P = (p' + p) / 2$, $q = p' - p$,
$r_\mu = 2 \varepsilon_{\mu \alpha \beta \gamma} q^\alpha P^\beta
\gamma^\gamma \gamma_5$ and $G_M (q^2 = 0) = \mu_N$ is the nucleon
total magnetic moment.  The corresponding proton and neutron magnetic
moments expressed in terms of the valence quark spins of (1) are~\cite{rf1}.
$$
	\mu_{p} = \mu_u \Delta u_V + \mu_d \Delta d_V + \mu_s \Delta s_V~,
\eqno(2a)
$$
$$
	\mu_n = \mu_d \Delta u_V + \mu_u \Delta d_V + \mu_s \Delta s_V~,
\eqno(2b)
$$
where $\mu_{p}$, $\mu_n = 2.793$, $-1.913$ are scaled to nucleon magnetons
$e  / 2 m_N$ and the corresponding quark magnetons are in the ratio
of quark charges and constituent quark masses as
$$
	\mu_u / \mu_d \approx e_u / e_d = -2~,\sep{.3}
	\mu_s / \mu_d = e_s m_d / e_d m_s \approx \smfrac{2}{3}~,
\eqno(2c)
$$
because in the latter relation one expects the inverse mass ratio to be 
$m_s/m_d\approx1.5$~\cite{rf1}.
Then substituting the quark valence spin values (1) and the first quark
magneton ratio of (2c) into the quark spin versions of the nucleon
magnetic moment valence expressions (2a, b), one obtains
the predicted ratio~\cite{rf1}.
$$
	\frac{\mu_{p}}{\mu_n} =
	\frac{\pip{\frac{\mu_u}{\mu_d} \frac{\Delta u_V}{\Delta d_V}}+1+0}
		{\frac{\Delta u_V}{\Delta d_V} + \frac{\mu_u}{\mu_d} + 0} =
	\frac{9}{-6} = -1.50~,
\eqno(2d)
$$
very close to the observed magnetic moment ratio~\cite{rf2} $\mu_{p} / 
\mu_n \approx -1.460$.

	Also subtracting or adding the nucleon magnetic moment relations
(2a,b) leads to quite reasonable values of the nonstrange $(NS)$
constituent quark mass ($m_u = m_d = \hat{m}$) when the valence
values in (1) are again used:
$$
	\mu_p - \mu_n = (\mu_u - \mu_d) (\Delta u_V - \Delta d_V)
	\approx 4.71 (e / 2 m_N)
	\eqno(2e)
$$
$$
	\mu_p + \mu_n = (\mu_u + \mu_d) ( \Delta u_V + \Delta d_V ) +
		2 \mu_s \Delta s_V \approx 0.88 (e / 2 m_N)
	\eqno(2f)
$$
Equations (2e,f) require $\hat{m} \approx 332$ MeV, 356 MeV, respectively.
Such an average NS constituent quark mass as 340 MeV can be obtained in 
many other ways as well~\cite{rf3} and is compatible with many data sets.

Furthermore, the magnetic moment quark valence expressions of the
$\Sigma^{+,-}$ baryons in analogy with eqs.\ (2a,b) for $\Sigma^+ =
uus$ and $\Sigma^- = dds$ are
$$
	\mu_{\Sigma^+} = \mu_u \Delta u_V + \mu_s \Delta d_V + \mu_d
		\Delta s_V
	\eqno(3a)
$$
$$
	\mu_{\Sigma^-} = \mu_d \Delta u_V + \mu_s \Delta d_V + \mu_u
		\Delta s_V~~.
	\eqno(3b)
$$
Then the measured difference of magnetic moments~\cite{rf2} 
$\mu_{\Sigma^+} \approx 2.458$, $\mu_{\Sigma^-} \approx -1.160$ in units 
of nuclear magnetons combined with the nucleon magnetic moment difference 
from (2a,b) both computed in the quark valence picture of (1) predicts the
magnetic moment ratio
$$
	\frac{\mu_{\Sigma^+} - \mu_{\Sigma^-}}
		{\mu_{p} - \mu_n} =
	\frac{\Delta u_V - \Delta s_V}
		{\Delta u_V - \Delta d_V} =
	\frac{\smfrac{4}{3}}
		{\smfrac{5}{3}} =
	0.80~.
\eqno(3c)
$$
This ratio (3c) is likewise close to the measured ratio 0.77.

	It is clear that the above quark valence picture cannot be the
whole story, however, because the nonstrange $\lambda_3$ isovector difference
of quark spins determined by neutron $\beta$ decay~\cite{rf2},
$$
	\Delta u - \Delta d = g_A =
	1.2670 \pm 0.0035 ~,
\eqno(4)
$$
is 25\% less than the valence value in (1); i.e.\ $\Delta u_V - 
\Delta d_V = \frac{5}{3}$. Moreover, the $\lambda_8$ component of the
nucleon axial-vector current found from the various hyperon 
semileptonic weak decays 
$$
	\Delta u + \Delta d - 2 \Delta s =
	g_A \pib{\frac{3f - d}{f+d}}_A \approx 0.58
\eqno(5)
$$
for the empirically determined~\cite{rf4} ratio $(d / f)_A \approx 1.74$
(or equivalently $(f/d)_A \approx 0.58$), is 40\% less than
the valence $\lambda_8$ prediction $\Delta u_V + \Delta d_V -
2 \Delta s_V = 1$.  However, these more dynamical quark-spin
difference SU(3) relations, (4) and (5) have more dynamical significance.
Adding and subtracting these quark spin relations (4) and (5) lead to 
two more (equivalent) quark spin SU(3) relations~\cite{rf5}
$$
	\Delta u - \Delta s \approx 0.92~,\sep{.3}
	\Delta d - \Delta s \approx -0.34~.
\eqno(6)
$$
For example, the ratio of the first difference in (6) to the $\lambda_3$ 
difference in (4) gives
$$
	\frac{\Delta u - \Delta s}
		{\Delta u - \Delta d} =
	\pib{ \frac{2f}{f+d}}_A \approx
	0.73~,
\eqno(7)
$$
instead of the ratio in (3c), which in turn 
predicts the sigma to nucleon magnetic
moment difference to be 0.73, also close to the experimental ratio
of 0.77~.

	However, these more dynamical quark spin differences (4)--(6) do
	a poor job when combined with the (static) nucleon magnetic
	moment difference from (2a,b), predicting instead
	$$
		\mu_p - \mu_n =
		(\mu_u - \mu_d) (\Delta u - \Delta d) \approx
		(e / 2m_d) g_A \approx
		4.71 (e / 2m_N)~,
		\eqno(8)
$$
or a smaller-than-expected constituent quark mass of $m_d \approx
253 ~\w{MeV}$.

\section*{II.  Dynamical QCD Models}
To apply  these four quark spin difference relations in (4), (5) and (6) 
to more dynamical problems, one must introduce a (chiral) model to set 
the overall quark spin scale.  In QCD the gluon spin $\Delta g$ must also 
be accounted for via the helicity sum rule resulting in the spin 
$\frac{1}{2}$ nucleon:
$$
	s_N = \smfrac{1}{2} = \smfrac{1}{2} \Delta \Sigma + \Delta g
	+ \Delta L_Z~,
\eqno(9a)
$$
where $\Delta L_Z$ is the orbital angular momentum of the constituent
quarks and $\Delta \Sigma = \Delta u + \Delta d + \Delta s$ is the
total quark spin of the nucleon ($\Delta \Sigma_V = 1$ in the valence 
picture of (1)).  The SLAC scaling experiments of 1970~\cite{rf6,rf5} 
suggest at a 1 GeV scale $- \Delta L_Z = k_\unka r_c \approx 1.3$ for
$k_\unka \approx 300 ~\w{MeV}$ and a proton transverse radius of
$r_c \approx 0.85 ~\w{fm}$ (the proton charge radius)\footnote{The 
fact that $\Delta g > 0$ (and $\Delta L_Z < 0$) is compatible with
finding more positive helicity gluons emitted by quarks inside a 
positive helicity proton.}. Then the
helicity sum rule (9a) at a 1 GeV scale reduces to
$$
	\smfrac{1}{2} \Delta \Sigma + \Delta g \approx 1.8~.
\eqno(9b)
$$

On the other hand~\cite{rf7}, $\Delta q \to \Delta q - (\alpha_s \Delta g) /
2\pi$ in QCD, but the gluon spin does not affect the SU(3) quark spin
difference relations (4)--(6) because $\Delta g$ is 
flavor-independent.  However, $\Delta g$ does enter spin 
singlet-dependent relations such as the EMC measurement~\cite{rf8} of the
polarized proton structure function integral at $q^2 \approx 10
~\w{GeV}^2$:
$$
	\int_{0}^{1} dx g_1^{p} (x) =
	\smfrac{1}{2} \pib{ \smfrac{4}{9} \Delta u +
					\smfrac{1}{9} \Delta d +
					\smfrac{1}{9} \Delta s}	-
	\smfrac{1}{3} \pip{ \frac{\alpha_s \Delta g}{2\pi}} \approx 0.126~.
\eqno(10)
$$
Here the combination $\alpha_s \Delta g$ is an approximate 
renormalization group invariant, as are the quark spins, which we
compute at a 1 GeV scale where the reduced helicity sum rule (9b)
containing $\Delta g$ is obtained, and it is known~\cite{rf9} that
$\alpha_s (1 ~\w{GeV}^2) \approx 0.50$.  Then substituting $\Delta g$
from (9b) together with $\alpha_s \approx 0.50$ into the EMC result
(10), one obtains a pure quark spin scale (including $U(3)$ singlet
components)
$$
	4 \Delta u + \Delta d + \Delta s + 0.24 \Delta \Sigma \approx 3.13~.
\eqno(11)
$$

	The solution of the quark spin difference relations (4)--(6) 
together with the EMC-deduced quark spin scale (11) leads to the
unique solution~\cite{rf5}
$$
	\Delta u = 0.87~,\sep{.3}
	\Delta d = -0.40~,\sep{.3}
	\Delta s = -0.05~,\sep{.3}
	\Delta \Sigma = 0.42~.
\eqno(12)
$$
Then $\Delta \Sigma = 0.42$ from (12) in turn fixes the gluon spin
in the reduced helicity sum rule (9b) to be $\Delta g \approx 1.59$
at a renormalization scale of 1 GeV.  This latter $\Delta g$ depends
only on the $\Delta L_Z$ as modeled from SLAC scaling data together
with the less model-dependent SU(3) quark spin difference relations
(4)--(6)~.

	An alternative QCD-dependent determination of the various quark
spins follows from exploiting the Bjorken sum rule~\cite{rf10} as it
pertains to QCD beyond leading order~\cite{rf11}.  Using the recent 
EMC~\cite{rf8} and SMC data~\cite{rf12} for proton structure functions 
and SLAC E142 data~\cite{rf13} for neutron structure functions, the 
Bjorken sum rule (BSR)
$$
	\int_{0}^{1} \pib{g_1^p (x) - g^n_1 (x)} dx =
	\smfrac{1}{6} g_A \pib{1 - \alpha_s (Q^2) /\pi }
\eqno(13)
$$
is shown to be valid to within 12\%~\cite{rf14}.  Given this BSR (13),
perturbative QCD corrections calculated up to ${\cal{O} }
\pib{\pip{\alpha_s / \pi}^3}$ and estimated through ${\cal{O}}
\pib{\pip{\alpha_s / \pi}^4}$ lead to the quark spin relations~\cite{rf11}
(including higher twist effects)
$$
	\begin{displaystyle}
		\begin{array}{ll}
			\Delta u = 0.85 \pm 0.03~,\sep{.3}&
			\Delta d = -0.41 \pm 0.03\\
			\Delta s = -0.08 \pm 0.03~,\sep{.3}&
			\Delta \Sigma = 0.37 \pm 0.07~.
		\end{array}
	\end{displaystyle}
\eqno(14)
$$
Also to order ${\cal{O}} \pib{\pip{\alpha_s / \pi}^4}$, ref.~\cite{rf11}
finds $\alpha_s (2.5 ~\w{GeV}^2) = 0.375 \stackrel{+0.062}{_{-0.081}}$
(compatible with $\alpha_s (1 ~\w{GeV}^2) \approx 0.50$ in ref.~\cite{rf9}),
corresponding to $\alpha_s (M_Z^2) = 0.122 \pm 0.007$ as found in the
PDG tables~\cite{rf2} due to averaging data of PEP/PETRA, TRISTAN, LEP,
SLC, CLEO.

	We stress that the BSR structure (but without the $\alpha_s$ QCD
factor in (13)) already follows by subtracting the polarized
neutron structure function analog of (10) but with $u 
\leftrightarrow d$ quarks from the proton version (10) along with
using the $\lambda_3$ isovector difference of quark spins
$\Delta u - \Delta d = g_A$ of eq.\ (4).  The flavor-independent
gluon spin factor $\alpha_s \Delta g$ in (10) then cancels out
in the BSR relation (13), yet the BSR plus the higher order
dynamical QCD renormalization scheme ``knows'' about the second
SU(3) $d/f$-type of quark spin difference (5) (or (6)) and knows 
about the singlet-type of helicity sum rule (HSR) (9b) coupled with the EMC 
proton polarized structure function scale (10) or the 
(singlet-included) quark spin U(3) scale (11).  We draw this
conclusion because of the close agreement between the SU(3)
quark spin differences (4)--(6) plus the dynamical HSR determination
of the EMC measurement (10) leading to the quark spins of eqs.\ (12).
The latter roughly numerically match the dynamically driven QCD analysis
of the BSR (13) resulting in the quark spins of eqs.\ (14).

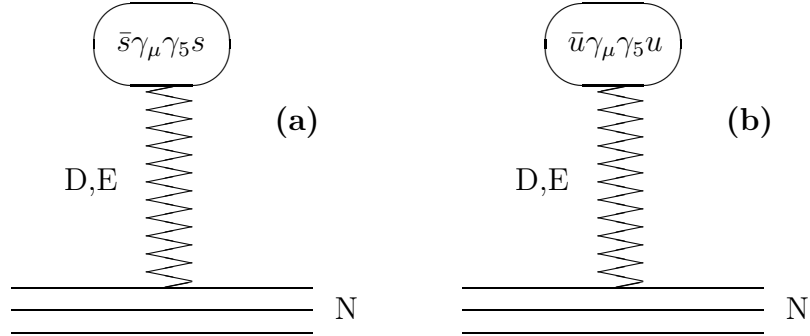
\begin{figure}[hbt] \centering
  \begin{picture}(135,65)

  \put(15,3){\line(1,0){40}}
  \put(15,6){\line(1,0){40}}
  \put(15,9){\line(1,0){40}}

  \put(35,9){\line(5,1){4}}
  \put(39,9.9){\line(-5,1){6}}
  \put(33,11.1){\line(5,1){6}}
  \put(39,12.3){\line(-5,1){6}}
  \put(33,13.5){\line(5,1){6}}
  \put(39,14.7){\line(-5,1){6}}
  \put(33,15.9){\line(5,1){6}}
  \put(39,17.1){\line(-5,1){6}}
  \put(33,18.3){\line(5,1){6}}
  \put(39,19.5){\line(-5,1){6}}
  \put(33,20.7){\line(5,1){6}}
  \put(39,21.9){\line(-5,1){6}}
  \put(33,23.1){\line(5,1){6}}
  \put(39,24.3){\line(-5,1){6}}
  \put(33,25.5){\line(5,1){6}}
  \put(39,26.7){\line(-5,1){6}}
  \put(33,27.9){\line(5,1){6}}
  \put(39,29.1){\line(-5,1){6}}
  \put(33,30.3){\line(5,1){6}}
  \put(39,31.5){\line(-5,1){6}}
  \put(33,32.7){\line(5,1){6}}
  \put(39,33.9){\line(-5,1){6}}
  \put(33,35.1){\line(5,1){4}}

  \put(35,41.4){\oval(18,11)}
  \put(29,40.5){$\bar{s}\gamma_{\mu}\gamma_{5}s$}
  \put(22,22){D,E}
  \put(58,4.8){N}
  \put(50,30){\bf (a)}

  \put(75,3){\line(1,0){40}}
  \put(75,6){\line(1,0){40}}
  \put(75,9){\line(1,0){40}}

  \put(95,9){\line(5,1){4}}
  \put(99,9.9){\line(-5,1){6}}
  \put(93,11.1){\line(5,1){6}}
  \put(99,12.3){\line(-5,1){6}}
  \put(93,13.5){\line(5,1){6}}
  \put(99,14.7){\line(-5,1){6}}
  \put(93,15.9){\line(5,1){6}}
  \put(99,17.1){\line(-5,1){6}}
  \put(93,18.3){\line(5,1){6}}
  \put(99,19.5){\line(-5,1){6}}
  \put(93,20.7){\line(5,1){6}}
  \put(99,21.9){\line(-5,1){6}}
  \put(93,23.1){\line(5,1){6}}
  \put(99,24.3){\line(-5,1){6}}
  \put(93,25.5){\line(5,1){6}}
  \put(99,26.7){\line(-5,1){6}}
  \put(93,27.9){\line(5,1){6}}
  \put(99,29.1){\line(-5,1){6}}
  \put(93,30.3){\line(5,1){6}}
  \put(99,31.5){\line(-5,1){6}}
  \put(93,32.7){\line(5,1){6}}
  \put(99,33.9){\line(-5,1){6}}
  \put(93,35.1){\line(5,1){4}}

  \put(95,41.4){\oval(18,11)}
  \put(89,40.5){$\bar{u}\gamma_{\mu}\gamma_{5}u$}
  \put(82,22){D,E}
  \put(118,4.8){N}
  \put(110,30){\bf (b)}

  \end{picture}

  \caption{\footnotesize (a) Strange-quark contamination of nucleons 
           via axial-vector $D$, $E$ meson tadpole graphs.  
           (b) Up-quark axial-vector content of nucleons via 
           axial-vector $D$, $E$ meson tadpole graphs.}

  \label{fig:one}
\end{figure}

\section*{III.  Dynamical Tadpole Leakage}

	A third scheme to estimate the quark axial-vector spins
$\Delta u$, $\Delta d$, $\Delta s$ which naturally links up to
the quark-valence picture of Sec.\ I is due to axial-vector tadpole
graphs.  For simplicity we shall label the 
axial-vector hadronic states $f_1 (1285)$ and $f_1 (1420)$ by their
early names, $D$ and $E$ respectively.  Then we use the nonstrange
$(NS)$ and strange $(S)$ hadronic axial-vector basis to write~\cite{rf15}
$$
	\rip{D} = \cos \phi_A \rip{A_{NS}} - \sin \phi_A \rip{A_S}~,\sep{.1}
	\rip{E} = \sin \phi_A \rip{A_{NS}} + \cos \phi_A \rip{A_S}~,
\eqno(15)
$$
where $\rip{A_{NS}} = \rip{\ov{u} u + \ov{d}d} / \sqrt{2}$ and
$\rip{A_S} = \rip{\ov{s}s}$~.  The axial-vector mixing angle in (15)
is determined by the present decay rate ratio --- the branching fraction 
for $D\rightarrow K\ov{K}\pi$ is 9\% and we take 50\% for 
$E\rightarrow K\ov{K}\pi$ since this mode together with
$E\rightarrow K\ov{K}^{\ast} +$ c.c. are both dominant~\cite{rf2}:
$$
	\frac{\Gamma_{DK\ov{K}\pi}}
		{\Gamma_{EK\ov{K}\pi}} \approx
	\tan^2 \phi_A \approx
	\frac{2.16 ~\w{MeV}}
		{27.75 ~\w{MeV}} \approx
	0.078~,\sep{.3}	\phi_A \approx 16\deg~.
\eqno(16)
$$

	Given eqs.\ (15) and (16), the quark spin $\Delta s$ is found from
Fig.~1a as~\mbox{\cite{rf16,rf5}}
$$
	\Delta s =
	- \pib{\frac{1}{m_D^2} - \frac{1}{m_E^2}} 
	\cos \phi_A 	\sin \phi_A 
	\trio{0}{\ov{s} \gamma_\mu \gamma_5 s}{A_S} 
	\left\langle A_{NS} N | N \right\rangle ^\mu~.
\eqno(17a)
$$
To link up with Sec.\ I, SU(2) symmetry is assumed conserved at the
lower vertex in Fig.~1a so that $\left\langle A_{NS} N | N 
\right\rangle ^\mu = 1 \cdot S^\mu$, where $S^\mu$ is the spin 1 
polarization four vector, along with $\trio{0}{\ov{s} \gamma_\mu
\gamma_5 s}{A_S} = S_\mu m_{A_S}^2$ (with $m_{A_S}^2 = m_D^2
\sin^2 \phi_A + m_E^2 \cos^2 \phi_A \approx 2 ~\w{GeV}^2)$.  Then
the tadpole equation (17a) requires
$$
	\Delta s =
	- \pib{ \frac{1}{m_D^2} - \frac{1}{m_E^2}}
	\cos 16\deg \sin 16\deg m_{A_S}^2 \approx -0.06~,
\eqno(17b)
$$
since~\cite{rf2} $m_D \approx 1282 ~\w{MeV}$ and $m_E \approx 1426 ~\w{MeV}$.

	Encouraged by the proximity of (17b) to our earlier determinations
in (12) and in (14) for $\Delta s$, we extend this analysis to
the tadpole graph of Fig.~1b in order to estimate the quark
spin $\Delta u$ in nucleons, leading to 
$$
	\Delta u = 
	\pib { \frac{\cos^2 \phi_A}
			{m_D^2}			+
		  \frac{\sin^2 \phi_A}
			{m_E^2}} 
	\trio{0}{\ov{u}\gamma_\mu \gamma_5 u}{A_{NS}} 
	\ipt{A_{NS} N}{N}^\mu~.
\eqno(18a)
$$
The first two terms in brackets in (18a) sum to 0.6 GeV$^{-2}$,
while the second factor is $\trio{0}{\ov{u} \gamma_\mu \gamma_5
u}{A_{NS}} = \frac{1}{\sqrt{2}} \trio{0}{\ov{u}\gamma_\mu \gamma_5
u}{\ov{u}u}$ and the latter up flavor matrix element is 
$\trio{0}{\ov{u} \gamma_u \gamma_5 u}{\ov{u}u} =
S_\mu \cdot 2 ~\w{GeV}^2$ in the SU(3) conserving limit as above
(17b).  Then the tadpole version of $\Delta u$ in (18a) becomes
$$
	\Delta u \approx 0.6 ~\w{GeV}^{-2} \cdot
	2 ~\w{GeV}^2 / \sqrt{2} \approx 0.85~,
\eqno(18b)
$$
again close to $\Delta u$ in (12) and in (14).

	Finally we compute the quark spin $\Delta d$ in this tadpole
picture driven again by $D$ and $E$ spin $1^+$ poles.  Since they
too simulate the axial-vector current, we can invoke the $\lambda_3$
axial current quark spin difference relation found from neutron
$\beta$ decay, $\Delta u - \Delta d = g_A$.  This latter relation was
used in part to fix the quark spins in (12) and it was also needed 
to derive the Bjorken sum rule (13) which in turn led to the quark
spins in (14).  So assuming the tadpole-driven $\Delta u \approx
0.85$ in (18b), the tadpole-driven $\Delta d$ is
$$
	\Delta d = \Delta u - g_A \approx 0.85 - 1.27 = -0.42~.
\eqno(19)
$$
Collecting the tadpole-driven quark spins of (17b), (18b) and (19), we
have
$$
	\Delta u \approx 0.85~,\sep{.3}
	\Delta d \approx -0.42~,\sep{.3}
	\Delta s \approx -0.06~,\sep{.3}
	\Delta \Sigma \approx 0.37~.
\eqno(20)
$$

\section*{IV.  Conclusion}
	We have reviewed the static SU(3) valence picture of the constituent
quark model while focusing on nucleon magnetic moments in Sec.\ I.
We slightly modified this picture in Sec.\ II in the context of
QCD dynamical models, first obtaining in eqs.\ (12) the strangeness
and total quark spins due to the helicity sum rule and EMC data
$$
	\Delta s = -0.05~,\sep{.3}
	\Delta \Sigma = \Delta u + \Delta d + \Delta s = 0.42~.
\eqno(21)
$$
Also in Sec.\ II we reviewed the Bjorken sum rule in QCD, resulting
in analog quark spins in eqs.\ (14),
$$
	\Delta s = -0.08 \pm 0.03~,\sep{.3}
	\Delta \Sigma = 0.37 \pm 0.07~.
\eqno(22)
$$
Finally in Sec.\ III we estimated the tadpole leakage due to $D$--$E$
mixing.  This led to eqs.\ (20) with strangeness and total quark spins
$$
	\Delta s \approx -0.06~,\sep{.3}
	\Delta \Sigma \approx 0.37~.
\eqno(23)
$$

	The consistent pattern of the above three dynamical schemes giving
(21), (22) and (23) makes us question whether such a compatible picture
is really a quark spin ``crisis''.  While it is true that the quark 
spin $\Delta s$ in (21)--(23) is a nonvanishingly small (and negative)
5\%--8\% of the valence values in (1), $\Sigma_V = 1$, $\Delta s_V = 0$,
the momentum distribution of strange quarks in nucleons was observed
by the Columbia-Chicago-Fermi-Rochester collaboration~\cite{rf17} to be a 
similar small 6\% fraction.  Moreover the SLAC scaling experiments of
the 1970s found that the overall momentum distribution was about 50\%
due to quarks and presumably 50\% due to QCD gluons.  In like manner,
the total quark spin $\Delta \Sigma$ in (21)--(23) is about 40\% and
implied gluon spin $\Delta g$ is about 60\% of the nucleon spin.  Even as a
crisis, the above quark spins always form an orderly pattern.

	The notion of crisis here asks why the quarks in (nonstrange) 
nucleons are not all nonstrange and why they do not always carry 100\% 
of the nucleon's spin, as the quark-valence picture requires.  We
have suggested in Sec.\ II that about 6\% of the nucleon's spin (and
also its momentum) are due to strange quarks.  Moreover, about 40\%
of the nucleon's total spin (and also its momentum) are due to quarks,
with the other 60\% presumably generated by QCD gluons.  An analogous
picture likewise emerges in Sec.\ III based on tadpole leakage of quark spins, 
assuming the (axial-vector) $D$ and $E$ tadpole propagators
couple to nucleons in an SU(3) and spin-conserving manner.
In the latter case, the empirical $D$--$E$ mixing can also be 
understood via OZI-violating quark-annihilation graphs coupling
to QCD gluons~\cite{rf15}.

	We conclude that in all three of the above quark spin
analyses in Secs.\ II and III QCD glue is needed to resolve the
quark strangeness and spin crisis in nucleons.  In a similar
fashion, one must be cautious when applying these dynamical quark
spin differences (4)--(6) to (static) magnetic moment problems such
as in eq.\ (8).  This too should not be considered a crisis, but
instead is a mixture of dynamical quark spins set in a static quark
framework.

\vspace{.5in}
\noindent
{\bf Acknowledgement}:  LB appreciates support from NSF.  MDS is grateful 
for 1997 summer support at TRIUMF.

\end{document}